\documentclass[twocolumn]{aastex6}
\usepackage{hyperref}
\usepackage{amsmath}

\newcommand{\J}{\mbox{J0815+0939}}
\newcommand{\B}{\mbox{B1839-04}}

\shorttitle{Bi-drifting subpulses of radio pulsar B1839-04}
\shortauthors{Szary et al.}

\begin{document}

\title{Single pulse modeling and the bi-drifting subpulses of radio pulsar B1839-04}
\author{Andrzej Szary\altaffilmark{1, 2}, Joeri van Leeuwen\altaffilmark{1,3}, Patrick Weltevrede\altaffilmark{4}, and Yogesh Maan\altaffilmark{1}}
\email{szary@astron.nl}

\altaffiltext{1}{ASTRON, Netherlands Institute for Radio Astronomy, Oude Hoogeveensedijk 4, 7991 PD, Dwingeloo, The Netherlands}
\altaffiltext{2}{Janusz Gil Institute of Astronomy, University of Zielona G\'ora, Lubuska 2, 65-265 Zielona G\'ora, Poland}
\altaffiltext{3}{Anton Pannekoek Institute for Astronomy, University of Amsterdam, Science Park 904, 1098 XH Amsterdam, Netherlands}
\altaffiltext{4}{Jodrell Bank Centre for Astrophysics, The University of Manchester, Alan Turing Building, Manchester, M13 9PL, United Kingdom} 

\begin{abstract}
We study the bi-drifting pulsar \mbox{B1839-04}, where the observed subpulse drift direction in the two leading pulse components is opposite from that in the two trailing components.
Such diametrically opposed apparent motions challenge our understanding of an underlying structure. 
We find that for the geometry spanned by the observer and the pulsar magnetic and rotation axes, the observed bi-drifting in \mbox{B1839-04} can be reproduced assuming a non-dipolar configuration of the surface magnetic field.
Acceptable solutions are found to either have relatively weak ($\sim 10^{12} \,{\rm G}$) or strong ($\sim 10^{14} \,{\rm G}$) surface magnetic fields.
Our single pulse modeling shows that a global electric potential variation at the polar cap that leads to a solid-body-like rotation of spark forming regions is favorable in reproducing the observed drift characteristics.  
This variation of the potential additionally ensures that the variability is identical in all pulse components resulting in the observed phase locking of subpulses.
Thorough and more general studies of pulsar geometry show that a low ratio of impact factor to opening angle ($\beta / \rho$) increases the likelihood of bi-drifting to be observed.
We thus conclude that bi-drifting is visible when our line of sight crosses close to the magnetic pole.
\end{abstract}

\section{Introduction}
The radio emission of pulsars is characterized by a sequence of highly periodic pulses. 
The single pulses consist of one or more components, the so-called subpulses, which, in many cases, exhibit systematic variation in phase or intensity or both. 
The drifting subpulses phenomenon was first observed by \cite{1968_Drake} and now we know around 120 pulsars which show this behaviour \citep{2006_Weltevrede, 2007_Weltevrede, 2016_Basu}.
The drifting is closely related to the processes responsible for radio emission, and as such is a good tool for testing physical mechanisms responsible for radio emission in pulsars. 
In the early years of pulsar research \cite{1975_Ruderman} proposed a carousel model to explain the drifting phenomenon. 
The model assumes that a group of localized discharges, the so-called sparks, circulate around the magnetic axis.
It turned out that this model can explain a variety of pulsar data. 
However, "the carousel model is yet an heuristic model that provides frustratingly little physical insight" \citep{2014_Rankin}.

In \citet{2012_Leeuwen}  the drift velocity of plasma relative to the neutron star is shown to depend on the variation of the electric potential over the polar cap.
Following this thinking, a modified version of the carousel model was proposed in \cite{2017_Szary}.
It was noted that the spark forming regions rotate not around the magnetic axis per se, but around the point of electric potential extremum at the polar cap: minimum in the pulsar case ($\mathbf{\Omega} \cdot \boldsymbol{\mu} < 0$) and maximum in the antipulsar case ($\mathbf{\Omega} \cdot \boldsymbol{\mu} > 0$).
Here $\mathbf{\Omega}$ is the pulsar angular velocity and $\boldsymbol{\mu}$ is the magnetic moment vector.
In this paper we consider a pulsar case with a net positive charge density at the polar cap.
Plasma columns produced in the inner acceleration region stream into the magnetosphere and produce the observed radio emission at much higher altitudes \citep[see e.g.,][]{2000_Melikidze, 2002_Mitra, 2003_Kijak}.

For pulsars with multi-component pulse profiles, subpulses generally drift in the same direction.
There are only three pulsars (J0815+0939, J1034-3224, and B1839-04) known to exhibit clear "bi-drifting", an effect where the drift direction of subpulses is different in different pulse profile components \citep[see, e.g.][and references therein]{2019_Basu}.
\cite{2017_Wright} proposed that an elliptical, tilted, eccentric motion of emitting regions around the magnetic axis can account for the observed bi-drifting subpulses in pulsars J0815+0939 and B1839-04.
\cite{2017_Szary} demonstrate that such a configuration can be obtained in the framework of the modified carousel model using the non-dipolar surface magnetic field in J0815+0939.

In this paper we introduce single-pulse modeling, which we use to explain the unusual drift feature reported for radio pulsar B1839-04 \citep{2016_Weltevrede}.
The paper is organized as follows.
In Section \ref{sec:obs} we re-analyse subpulse modulation of PSR \B~ and quantify drift rates of pulse components, while in Section \ref{sec:model} we introduce the single pulse modeling procedure.
In Section \ref{sec:results} we compare our measurements with the modeled sample. 
The discussion is in Section \ref{sec:discussion}, followed by conclusions presented in Section \ref{sec:conclusions}.

\section{Observational properties of \mbox{PSR \B}} \label{sec:obs}

\begin{figure}[!t]
  \centerline{
  \includegraphics[width=8.1cm]{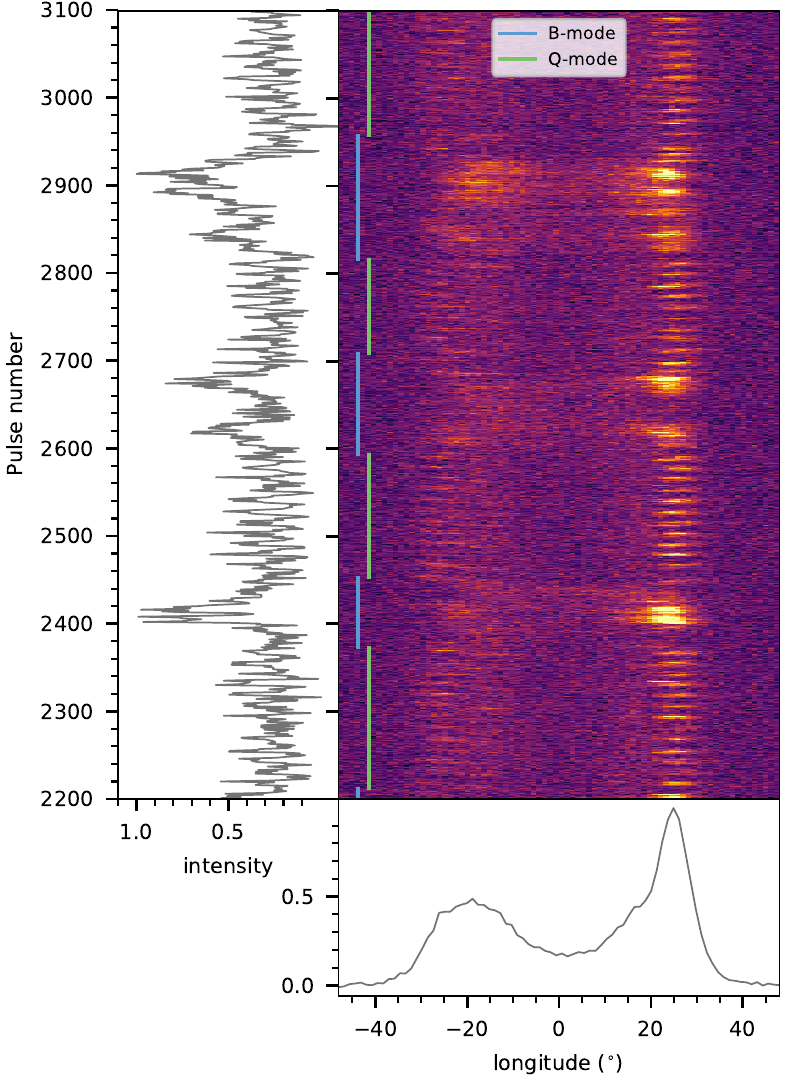}
  }
    \caption{The single pulses in \B.
    Intensity plot of 900 pulses is shown in {\it the main panel}.
    {\it The bottom} panel shows the integrated profile for this sequence.
    Variations in the single-pulse intensity are plotted in {\it the left panel}.
    The green and blue solid lines correspond to the Q-mode and B-mode, respectively. 
    \label{fig:single_pulses}
    }
\end{figure}

Pulsar \B~ was discovered with the Lovell telescope at Jodrell Bank \citep{1986_Clifton}.
The spin properties of the pulsar are not in any way extraordinary, as it is a slow rotator, $P=1.84\,{\rm s}$, with a relatively small period derivative, $\dot{P}=5.1\times 10^{-16} \,{\rm s/s }$, \citep[][]{2004_Hobbs}.
The derived characteristic age, $\tau_{\rm c}=57 \,{\rm Myr}$, and relatively small rate of loss of rotational energy, $\dot{E}=3.2\times 10^{30}\,{\rm erg\,s^{-1}}$ place it somewhat close to the graveyard region in the $P-\dot{P}$ diagram \citep{1993_Chen}.
Throughout this paper, we will use these rotational properties in our simulations.

\begin{figure}[!t]
  \centerline{
  \includegraphics[width=8.1cm]{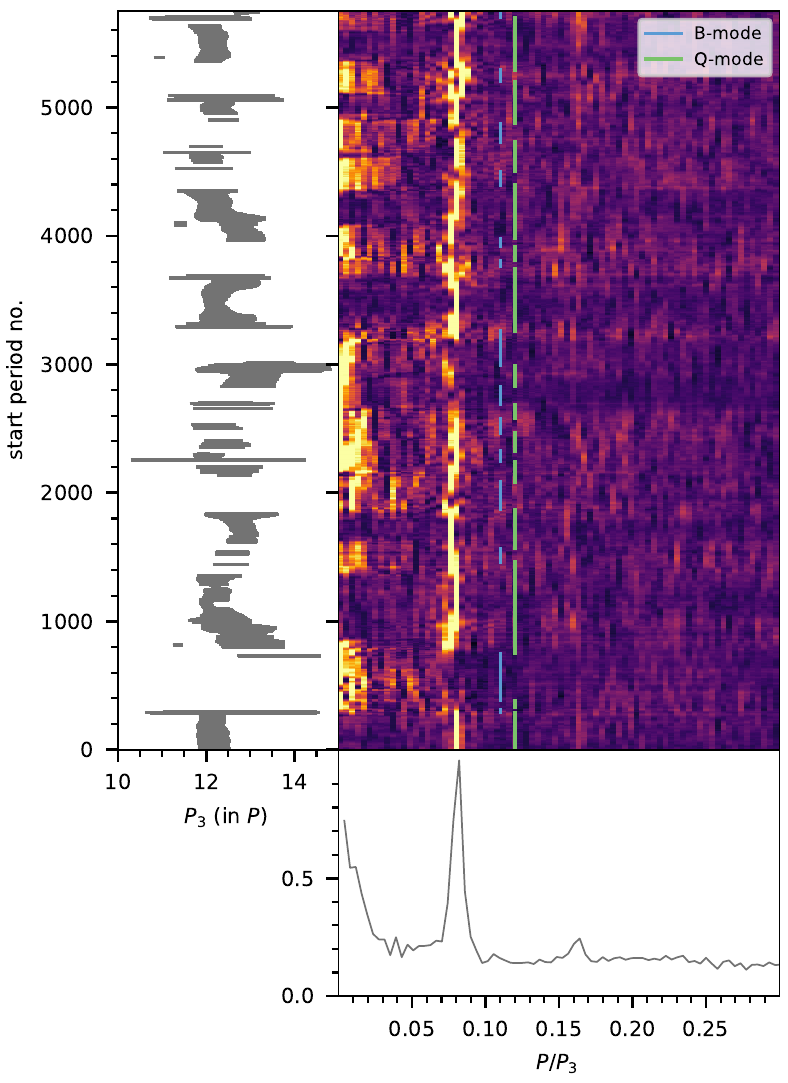}
  }
    \caption{Variation of the Longitude Resolved Fluctuation spectra (LRFS) as a function of the start period. The LRFS is determined for 256 consecutive single pulses. The starting point is shifted by one period and the process is repeated until the end of the observing session. The measured $P_3$ value  is shown in {\it the left panel}, while {\it the bottom panel} shows the time average LRFS.
    \label{fig:p3_evolution}
    }
\end{figure}

\begin{figure*}[!ht]
  \centerline{
  \includegraphics[width=15.7cm]{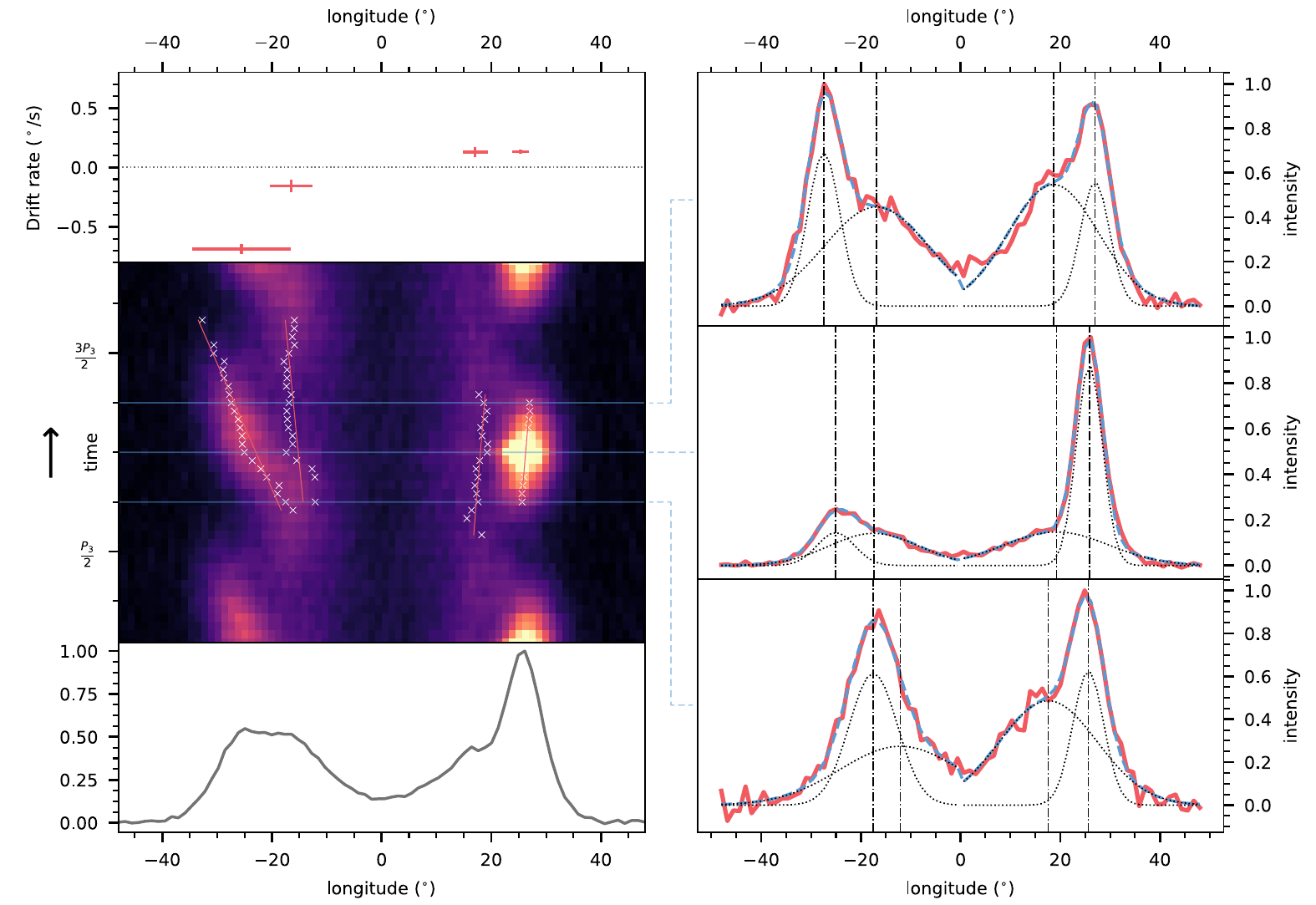}
  }
    \caption{{\it The middle left panel} shows the average drift band profile, obtained by folding the single-pulse signal in the Q-mode with a typical $P_3=12.3 \, P$.
    The white crosses correspond to the component positions and the red lines show the linear regression. 
    The blue horizontal lines mark the times for which we show the component detection procedure in the right panels. 
    {\it The top left panel} shows the measured drift rate, while {\it the bottom left panel} shows the integrated pulse profile.
    {\it In the right panels}, the red solid lines show horizontal cuts throughout the drift band profile, the blue dashed lines show the best-fitting analytic function which reproduces the profile (with the individual components shown with the dotted lines), finally the dot-dashed lines show the fitted longitudes of all four components.
    \label{fig:p3fold}
    }
\end{figure*}

\subsection{Observation}
The \B~ data used in this paper were recorded with the Westerbork Synthesis Radio Telescope on 2005 October 31. 
The pulsar was observed around a center frequency of $1380\,{\rm MHz}$ with a bandwidth of $80\,{\rm MHz}$ at $819 \mu {\rm s}$ time resolution.
The recorded 6008 pulses were dedispersed at 196\,pc cm$^{-3}$.
A detailed explanation of the process of recording and forming the single pulse data can be found in \cite{2006_Weltevrede}.

In Figure \ref{fig:single_pulses} we show a sequence of single pulses with clearly visible two distinct emission modes, the bright mode (B-mode) and the quiet mode (Q-mode).
The modes separation was made by eye using the \mbox{PSRSALSA}\footnote{A Suite of ALgorithms for Statistical Analysis of pulsar data. The latest version and a tutorial can be downloaded from https://github.com/weltevrede/psrsalsa} package  \citep[see Section 4.2 in][for more details]{2016_Weltevrede}.

\subsection{Drift Characteristics}

    \begin{table*}[]
        \caption{Drift parameters of PSR \B }
        \begin{center} 
            \begin{tabular}{lcccc}
                \hline
                \hline
                Component: & I & II & III& IV \\
                Longitude: $(^{\circ})$ & 
$-26$ &
$-16$ &
$17$ &
$25$ \\
                Drift band width: $(^{\circ})$ & 
$18$ &
$8$ &
$4$ &
$2$ \\
                Drift rate: $\left ( ^{\circ} / s \right )$  &
$ -0.69 \pm  0.04$ &
$ -0.16 \pm  0.05$ &
$ 0.13 \pm  0.04$ &
$ 0.13 \pm  0.01$
\\
                \hline
            \end{tabular}
        \end{center}
        \label{tab:drift}
        {\bf Notes.} The standard errors of linear regression are quoted.
    \end{table*}

To study the drift characteristics in \B~ we use the Longitude Resolved Fluctuation Spectra \citep[LRFS,][]{1970_Backer}.
The computation of the LRFS involves discrete Fourier transforms of consecutive pulses along each longitude, allowing to explore how the drifting feature varies across the pulse window.
    
For the entire observing session we calculate the LRFS for 256 single pulses shifting the starting point by one period \citep{2009_Serylak}.
The repetition time of the drift pattern, $P_3$, is determined by fitting a Gaussian near the major peak in the fluctuation spectra. 
In Figure \ref{fig:p3_evolution} we show the variation of the LRFS as a function of the start period.
Two things stand out from the figure.
First, the periodic subpulse modulation is present only during the Q-mode (see, for instance, start period number from 0 to 250), while in the B-mode the fluctuation spectra are dominated by a red-noise-like component (see, for instance, start period number from 500 to 700).
Second, the transition between the modes is connected with a change in $P_3$.
The most notable example of such behavior is for a start period between 800 and 1100, when pulsar enters the Q-mode, where the repetition time of the drift pattern seems to gradually decrease.

Since the periodic subpulse modulation is present only during the Q-mode \citep[see also][]{2016_Weltevrede}, hereafter, we use only this mode to analyze drift features in \B.
Performing the LRFS analysis of the first 416 single pulses in the sequence results in $P_3=12.3 \pm 0.2 P$. 
In order to determine the drift rate of each component we analyze the average drift band shape obtained by folding the data with the period $P_3$.
Since the $P_3$ is variable throughout the observation (see Figure \ref{fig:p3_evolution}), we use PSRSALSA which detects and compensates fluctuations in $P_3$ to fold the data \citep[see Section 4.7 in ][for more details]{2016_Weltevrede}.
Figure \ref{fig:p3fold} shows the folded drift band in the Q-mode.
As all pulse components cannot be well resolved in the integrated pulse profile, we fit a sum of two Gaussian to each half of the on-pulse region for different phases in the modulation cycle (see the right panels in Figure \ref{fig:p3fold}).
Then, the paths that these subpulses trace in time were fit with straight lines (see the middle left panel in Figure \ref{fig:p3fold}). 
The resulting drift rates are shown in the top panel of Figure~\ref{fig:p3fold} and are listed in Table~\ref{tab:drift}.

\section{The model} \label{sec:model}

\subsection{Theoretical background} \label{sec:theoretical}

In our studies we adopt the approach of \cite{1975_Ruderman}, where localized discharges of small regions over the polar cap produce plasma columns that stream into the magnetosphere, where they produce the observed radio emission.
As shown in \cite{2012_Leeuwen} the drift velocity depends on the variation of the accelerating potential across the polar cap.
This idea was further developed in \cite{2017_Szary}, where it was shown that circulation of spark forming regions is governed by the global variation of electric potential across the polar cap.

The modified carousel model stems from the insight that plasma generated in the spark forming regions rotates around local potential extrema, while plasma in regions between sparks rotates around global potential extremum at the polar cap (see Figure 8 in \citealt{2017_Szary}).

The electrodynamics governing the plasma generation in the inner acceleration region is still an unsolved problem \citep{2016_Melrose}.
In general in pulsar magnetosphere there are two sources of electric field: an inductive electric field, due to time-changing magnetic field as pulsar rotates and a potential field, due to the charge density in the magnetosphere.
In this work, we use an oblique geometry ($\mathbf{\Omega} \cdot \mathbf{B} < 0$) to explain the drift characteristics in \mbox{B1839-04}.
Thus, the co-rotation electric field has both inductive (divergence-free) and potential (curl-free) components \citep{1967_Melrose}.
The value of an inductive electric field is the same as in the vacuum model \citep[see, e.g.,][]{1967_Pacini, 1968_Pacini}, while the source of a potential field is the co-rotation charge density.
Therefore, the lack of charges in the polar cap region influences only a potential component of electric field. 
As a consequence, to find the drift direction in an oblique rotator we can use the same approach as for an non-oblique rotator (see Appendix \ref{app} for more details).

\subsection{Simulation setup}

Following the approach presented in \cite{2017_Szary} the magnetic field is modeled as a superposition of the global dipole field and crust-anchored small scale magnetic anomalies \citep{2002_Gil}.
This configuration produces the non-dipolar magnetic field at the surface that is thought to underlie such observed features as changing profiles, small x-ray-emitting polar caps, and emission from aligned rotators. 
Already at a relatively short distance from the star the global dipole starts to dominate and the configuration of the magnetic field is indistinguishable from a pure dipole  \citep[see Figure 3 in][]{2015_Szary}.
How our line of sight cuts through this magnetic field depends on the magnetic axis inclination angle, $\alpha$, and the impact parameter, $\beta$, which is the closest approach the line of sight makes to the magnetic axis.  
In our simulations we use oblique geometries, with $\alpha \in (147, 177)$ to explain radio characteristics in \B~ (see Section \ref{sec:geometry} for more details).
The simulation setup consists of the following steps: (I) adding one crust-anchored magnetic anomaly with random position and strength to the global dipole, resulting in a non-dipolar surface magnetic field  
(II) finding the actual location of the polar cap by modeling the last open field lines from the emission region down to the stellar surface, (III) finding the imprint of the line of sight on the polar cap by modeling the open field lines connected to the line of sight.
Hereafter, the imprint of the observer's line of sight on the polar cap, which marks the region where the plasma responsible for radio emission is generated, is called the plasma line.

\begin{figure}[t]
  \centerline{
  \includegraphics[width=6.3cm]{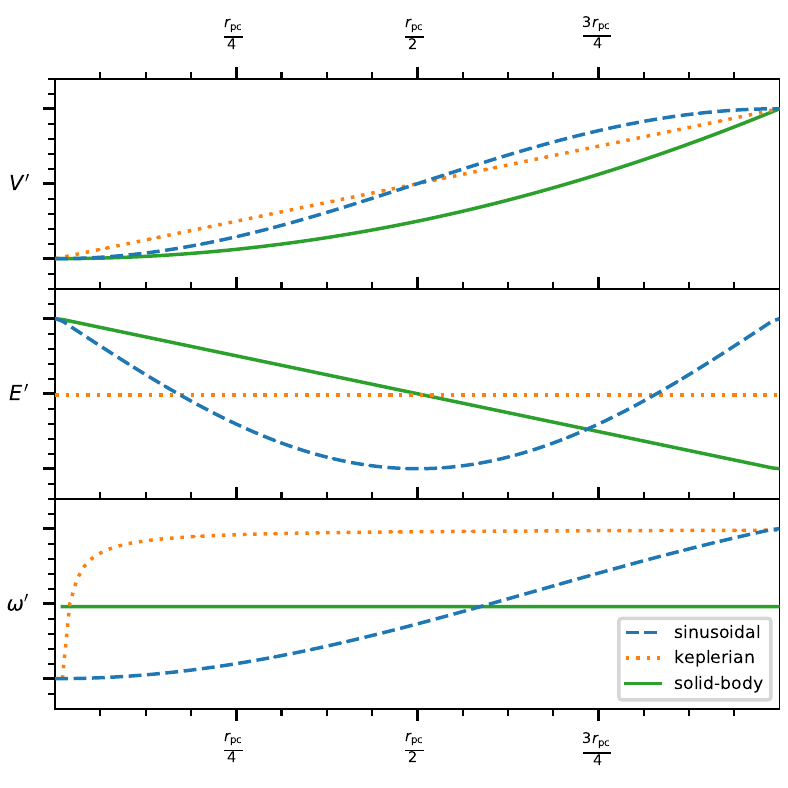}
  }
    \caption{Electric potential  {\it (the top panel)}, the electric field {\it (the middle panel)}, and the angular velocity of spark forming regions {\it (the bottom panel)} across the polar cap. The blue dashed line, the orange dotted line and the green solid line in the top panel correspond, respectively, to Equations 1, 2 and 3.
    \label{fig:potentials}
    }
\end{figure}


\subsection{Sparks rotation}

We use the modified carousel model \citep{2017_Szary}, where the circulation of spark forming regions is governed by the global variation of electric potential across the polar cap.
We examine several possible variations of the global potential at the polar cap in order to accurately describe and understand the observed single-pulse behavior in PSR \B:
\begin{displaymath}
    V^{\prime} (r, r_{\rm b}) \propto \left \{ \begin{array}{l} -  \left | \frac{V_0}{2} \right | \cos \left ( \frac{\pi r}{r_{\rm b}} \right ) \hspace{1.7cm} (1)\\ 
    |V_0| \left (r / r_{\rm b} \right ) \hspace{2.55cm} (2) \\ 
    |V_0| \left (r / r_{\rm b} \right )^2 \hspace{2.4cm} (3)\end{array} \right . ,
\end{displaymath}
\setcounter{equation}{3}
where $V_0$ is the potential amplitude, $r$ is the distance to the location of the potential minimum, and $r_{\rm b}$ is the radial distance from the location of the potential minimum to the polar cap boundary.
Note that for perfectly circular polar caps $r_{\rm b}$ is constant and equal to the polar cap radius (assuming potential minimum in the center of the polar cap). In general, however, $r_{\rm b}$ depends on the shape of the polar cap, and varies over it.
The electric potential described by Equation 1 corresponds to the sinusoidal variation used in \cite{2017_Szary};
The potential from Equations 2 and 3 results in keplerian-like rotation ($\omega^{\prime} \propto r^{-1}$) and solid-body-like rotation ($\omega^{\prime} ={\rm const.}$), respectively.
Where $\omega^{\prime}$ is the angular velocity of sparks in the co-rotating frame of reference.

Having defined the electric potential in the co-rotating frame, $V^{\prime}$, we can calculate the electric field perpendicular to the magnetic field $\widetilde{\bf E}^{^{\prime}}_{\perp}=-\nabla V^{\prime}$, and finally the drift velocity of spark forming regions 

\begin{equation}
{\bf v}' = c(\widetilde{\bf  E}^{^{\prime}}_{\perp} \times {\bf B})B^{-2}.
\label{eq:vdr}
\end{equation}
Here $\bf B$ is the magnetic field. 
Figure \ref{fig:potentials} shows the electric field {\it (the middle panel)} and the angular velocity of spark forming regions, \mbox{$\omega^{\prime} = v^{\prime} / r_{\rm c}$} {\it (the bottom panel)}, for different potential variations ({\it the top panel}).

\begin{figure}[tb]
  \includegraphics[width=8.0cm]{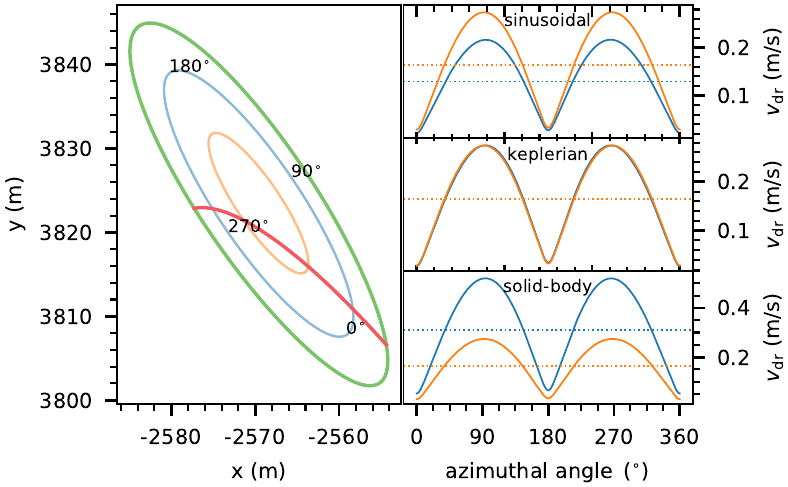}
    \caption{Modeled tracks of spark forming regions over the polar cap ({\it the left panel}) and the drift velocity of sparks ({\it the right panels}) for a particular configuration of magnetic field and pulsar geometry (see text for details).
    In {\it the left panel}, the green circle is the rim of the open field line region at the surface, the red line represents the plasma line (see text for description), the blue and orange lines represent the spark tracks. In {\it the right panels}, each corresponding to a different assumed potential across the polar cap, the orange and blue lines correspond to the drift velocity at, respectively, the inner and outer tracks as a function of the azimuthal angle of the spark relative to the centre of the polar cap. The dotted lines correspond to the mean drift velocity.
    \label{fig:tracks}
    }
\end{figure}

\subsection{Drift characteristics} \label{sec:drift_characteristics}

\begin{figure}[t]
  \includegraphics[width=8.0cm]{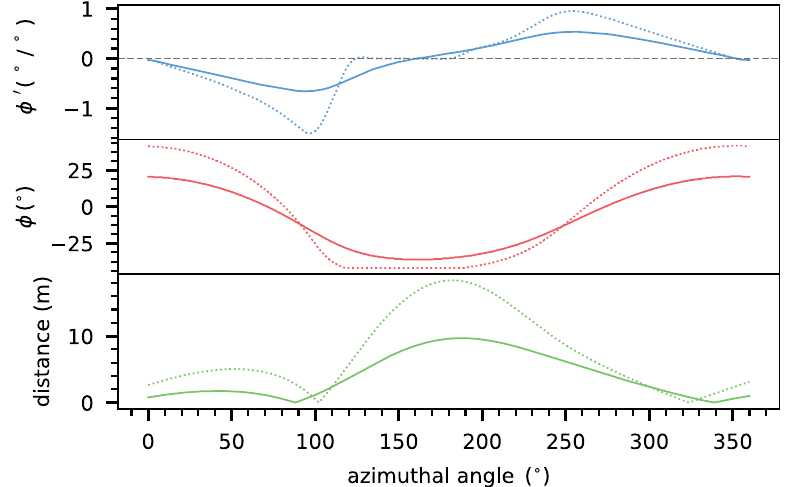}
    \caption{distance to the plasma line ({\it the bottom panel}), the simulated subpulse longitude ({\it the middle panel}) and the longitude derivative of an azimuthal angle ({\it the top panel}) as a function of an azimuthal angle of spark relative to the centre of the polar cap.
    the solid and dotted lines correspond to the inner and outer tracks, respectively.
    \label{fig:sparks_evolution}
    }
\end{figure}

In this paper we present calculations of the drift characteristics using a method that is an improvement over \citet{2017_Szary}.
Where that work modeled the bulk motions of the plasma in the polar cap, the present research models and follows the \emph{individual} sparks in that bulk motion.
We track the spark motions on the polar cap, translate these to the pulse longitudes of the resulting subpulses that  would be observed, and model the final pattern single pulses.
In this section we present results for a pulsar with $\alpha=170^{\circ}$, $\beta=1.2^{\circ}$ (see Section \ref{sec:geometry} for geometry considerations) and with a crust-anchored anomaly of 7\% of the dipole field strength, located at \mbox{($0.95R_{\rm NS}$, $31.18^{\circ}$, $123.90^{\circ}$)}.
Once the configuration of the magnetic field and the viewing geometry are set, the drift characteristics are modeled in the following steps:

(I) First, for a given number of profile components at a certain separation, the tracks of spark forming regions are found by deriving the propagation of sparks from the electric potential.
Figure \ref{fig:tracks} shows the modeled tracks (the left panel) and the drift velocity of sparks for the considered electric potential variations (the right panels).
The drift velocity is calculated assuming 11 sparks in the inner track and $P_3=23\,{\rm s}$, resulting in the mean velocity at the inner track $v_{\rm dr}=0.16 \, {\rm m / s}$.
As the sparks on the blue track cross the plasma line (the imprint of the line of sight; red in Figure \ref{fig:tracks}), they are responsible for the formation of first and fourth components in the profile, while on the orange track sparks are responsible for the formation of second and third components.
Note that for all three different descriptions of potential variation the spark tracks are the same, however, the drift velocity differs significantly.
During spark propagation we record its distance to the plasma line and the subpulse longitude (see Figure \ref{fig:sparks_evolution}).  
The subpulse longitude is the rotational phase derived from the position of a point at the plasma line which is nearest to the spark.

(II) The second step consist of finding the spark size which best reproduces the drift characteristics of PSR \B.
The spark size determines the largest distance from the plasma line at which the subpulses are still observed.
Since the derivative of subpulse longitude changes with distance from the plasma line (see the top panel in Figure \ref{fig:sparks_evolution}), the spark size affects the average drift rate.
A range of sizes is considered depending on the actual polar cap size.
For every spark size the drift band width for each component is calculated based on the subpulse rotational longitude.
Then, the simulated widths of drift bands are compared with the observed widths in the drift band profile (see the middle left panel in Figure \ref{fig:p3fold}), \mbox{$C = \sum_{n=1}^{4} \left( C_n^{\rm obs} -  C_n^{\rm sim} \right)^2 $}, where  $C_n^{\rm obs}$ is the measured drift band width of the {\it n-th} component (see Table \ref{tab:drift}), and $C_n^{\rm sim}$ is the simulated drift band width of the {\it n-th} component.

\begin{figure}[t]
  \centerline{\includegraphics[width=8.0cm]{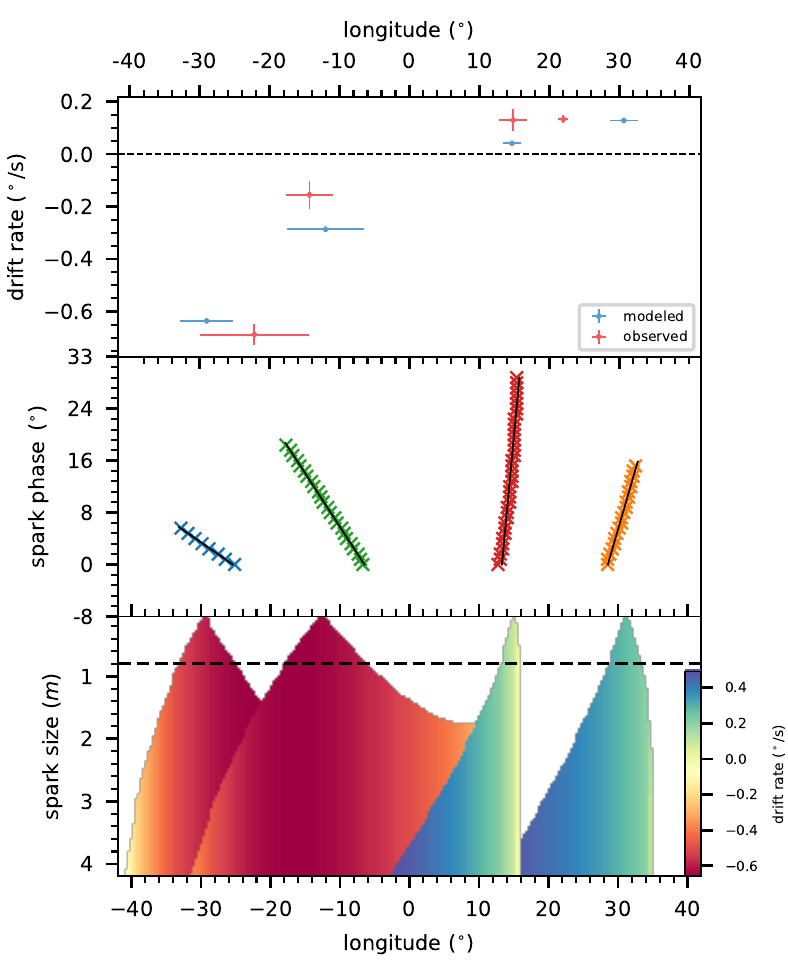}}
    \caption{
    \label{fig:drift_characteristics}
    Drift band widths and drift rates (shown with the colormap) for a range of spark sizes ({\it the bottom panel}).
    The simulated subpulse longitudes ({\it the middle panel}), and the drift rates ({\it the top panel}) for the spark size $0.85\,{\rm m}$ indicated with the dashed line in the lower panel and corresponding to the lowest value of $C$ (see text for more details).
    }
\end{figure}

(III) Finally, the drift rate of each component is calculated by fitting a straight line to the simulated subpulse longitudes (see Figure \ref{fig:drift_characteristics}).
We quantify the performance of the model by comparing the mean longitude variation of the simulated components with the measured drift rates, \mbox{$R = \sum_{n=1}^{4} \left( R_n^{\rm obs} -  R_n^{\rm sim} \right)^2$}, where  $R_n^{\rm obs}$ is the measured drift rate of the {\it n-th} component, and $R_n^{\rm sim}$ is the simulated drift rate of the {\it n-th} component.

\begin{figure*}[tb]
  \centerline{\includegraphics[width=15.9cm]{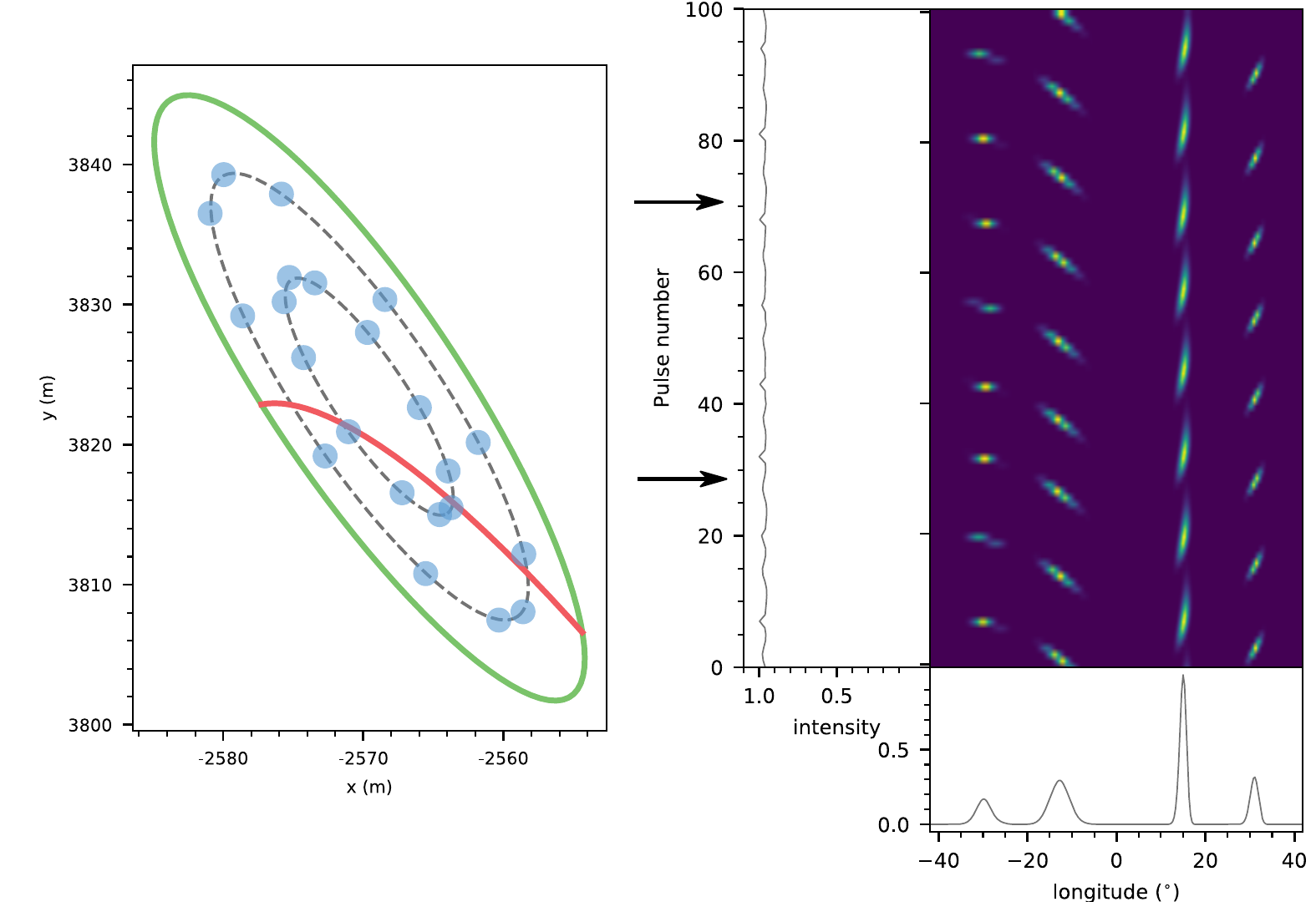}}
    \caption{Simulation setup of single pulse modeling ({\it the left panel}) and the modeled single pulses ({\it the right panels}) for a particular configuration of surface magnetic field and pulsar geometry (see text for details).
    The grey dashed lines in {\it the left panel} correspond to spark tracks and the blue circles represent sparks location and size.
    For description of {\it the right  panels}, see Figure \ref{fig:single_pulses}.
    \label{fig:modeled_pulses}
    }
\end{figure*}

\begin{table*}[t]
\centering
\caption{Number of realizations showing the bi-drifting behaviour (with $R < 0.1$).}
\label{tab:potential}
\begin{tabular}{ccccccc}
 \hline
 \hline
 \multicolumn{2}{c}{Geometry} & Emission height & Pulse width & \multicolumn{3}{c}{Bi-drifting probability $(\%)$} \\
$\alpha \left ( ^{\circ} \right)$  & $\beta \left ( ^{\circ} \right)$ & $D \,  ({\rm in} \, {\rm km})$ &  $W \left ( ^{\circ} \right) $ &sinusoidal & keplerian & solid-body \\
 \hline
147.0 & 3.5 & 3579 & 67.2 & 0.07 & 0.04 & 1.88 \\ 
150.3 & 3.3 & 2951 & 67.9 & 0.05 & 0.13 & 1.62 \\ 
153.7 & 3.0 & 2368 & 67.2 & 0.06 & 0.14   & 1.71 \\ 
157.0 & 2.6 & 1837 & 67.2 & 0.05 & 0.09   & 1.80 \\ 
160.3 & 2.3 & 1363 & 67.9 & 0.02 & 0.14   & 1.50 \\ 
163.7 & 1.9 & 950 & 67.1 & 0.02 & 0.13   & 1.27 \\ 
167.0 & 1.5 &  608 & 67.1 & 0.06 & 0.16 & 1.53 \\ 
170.3 & 1.2 &  338 & 67.0  & 0.01 & 0.08 & 1.40 \\ 
173.7 & 0.8 &  147 & 67.6 & 0.01 & 0.11   & 1.58 \\ 
177.0 & 0.4 &   33 & 68.1 & 0.04 & 0.11   & 1.23 \\ 
 \hline
\end{tabular}
\end{table*}

\subsection{Single pulses}
The parameters of single-pulse modeling are: $C_{\rm pos}$ - the components' positions in the profile expressed in phase, $S_{\rm i}$ - the number of spark forming regions at the inner track, $S_{\rm spp}$ - the number of simulation steps per pulsar period ($P$), $\sigma_{\rm S}$ - the spark width in meters (Gaussian standard deviation), and the repetition time of the subpulse pattern, $P_3$.
Note that the number of sparks at the outer tracks depends on the used electric potential variation (see Section \ref{sec:phase-locking} for more details).
The modeling of single pulses consist of the following steps.\\
(I) For given positions of components in a profile, the spark tracks at the polar cap are calculated.\\
(II) The tracks are filled with a given number of sparks. The distance between sparks is determined from the drift velocity along the track, such that the modeled $P_3$ value is the same between every spark. \\
(III) Next the evolution of spark positions is calculated with time resolution $dt= P / S_{\rm spp}$. \\
(IV) In every step, the intensity across the plasma line is calculated using 2D Gaussian functions, based on the sparks' location and width $\sigma_{\rm S}$.\\
The left panel in Figure \ref{fig:modeled_pulses} shows the simulation setup with parameters $C_{\rm pos}=[0.15, 0.35]$, $S_{\rm i}=11$, $S_{\rm spp}=20$, $\sigma_{\rm S}=0.28{\rm m}$ (which corresponds to the spark size $\sim 0.85\,{\rm m}$). 
In the calculation a signal from the radio pulsar is generated.
The single pulses are constructed by selecting samples with $P$ interval from the signal.
In the right panel of Figure \ref{fig:modeled_pulses} we show the modeled single pulses obtained using a potential variation described with Equation 3.
Note that in the proposed procedure the value of $V_0$ is set such that for a given number of spark forming regions at the inner track the value of $P_3$ is correct.


\section{Results} \label{sec:results}

\subsection{Phase locking}
\label{sec:phase-locking}

It was shown in \cite{2016_Weltevrede} that when the modulation cycle slows down in the leading components (I, II), it is equally slow in the trailing components (III, IV). 
Such phase locking of modulation patterns, resulting in identical $P_3$ values for all profile components, is very common in all pulsars \citep[see e.g.][]{2006_Weltevrede, 2007_Weltevrede, 2016_Basu}.
Furthermore, it strongly suggests that the modulation in all components has a single physical origin.

As noted in Section \ref{sec:drift_characteristics}, the tracks of spark-forming regions do not depend on the potential variation.
However, the drift velocity on each track can vary significantly (see the right panels in Figure \ref{fig:tracks}).
Therefore, in order to model the observed phase-locking of profile components, the number of sparks at the inner and outer tracks has to take into account this difference.
For the configuration presented in the left panel of Figure \ref{fig:modeled_pulses} the lengths of the inner and outer tracks are, respectively, $l_{\rm i} = 41.5 \,{\rm m}$ and $l_{\rm o}= 78.8 \,{\rm m}$.
The mean drift velocity at the inner track can be calculated using the number of spark forming regions, $S_{\rm i} = 11$, and $P_3=23\,{\rm s}$, resulting in $v_{\rm i}=0.16 \, {\rm m / s}$.
Next the mean drift velocities at the outer tracks are $v_{\rm o,1}=0.13\, {\rm m / s}$, $v_{\rm o,2}=0.16 \, {\rm m / s}$, $v_{\rm o,3}=0.31 \, {\rm m / s}$, respectively, for the potential variation described by Equations 1, 2, and 3.
To ensure the same value of $P_3$ at each track, the number of spark forming regions at the outer track can be calculated as $S_{o} = S_{\rm i} (l_{\rm o} / l_{\rm i})(v_{\rm i} / v_{\rm o})$, resulting in $S_{\rm o, 1}=26.5$, $S_{\rm o, 2}=20.9$, $S_{\rm o, 3}=11$.
The non-integer number of spark forming regions for cases with potential variation described by Equations 1 and 2 means that for defined spark tracks (i.e. a given component separation) the $P_3$ will vary at each track, and thus do not meet the observation requirement of phase-locking.
The phase-locking over a period of years implies that solid-body-like rotation (Equation 3), i.e. the same number of sparks on each track, is the most realistic assumption.

\begin{figure*}[!ht]
  \centering
  \includegraphics[width=16.5cm]{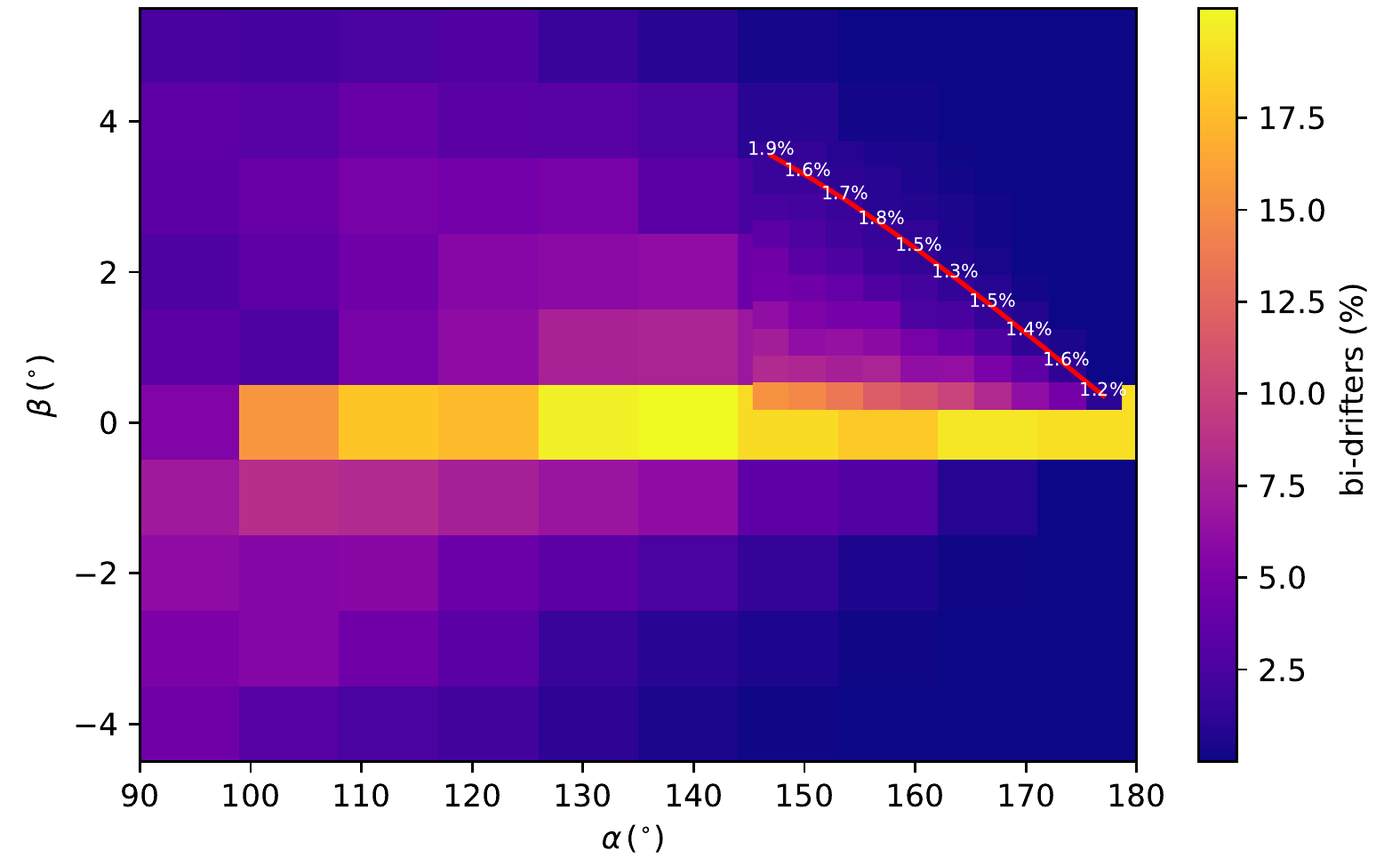}
    \caption{Fraction of pulsars showing bi-drifting (with $R<0.1$) for random configuration of surface magnetic field.
    \label{fig:fraction_grid}
    }
\end{figure*}

\begin{figure*}[!ht]
  \includegraphics[width=8.5cm]{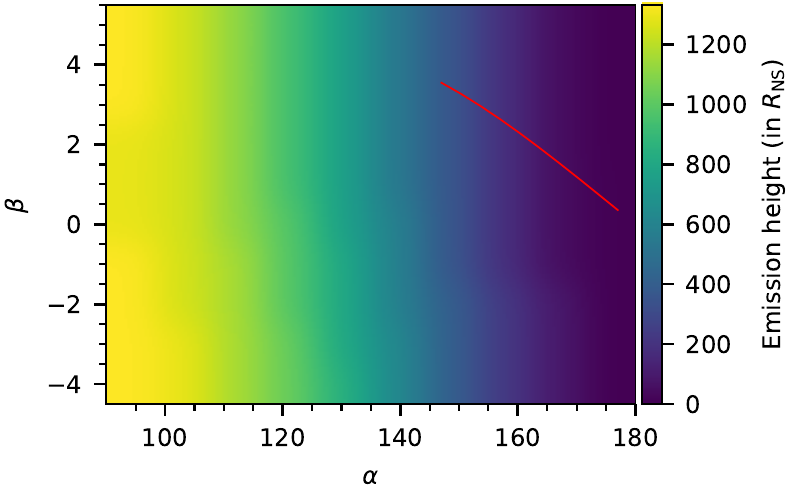} \hspace{1cm}
  \includegraphics[width=8.5cm]{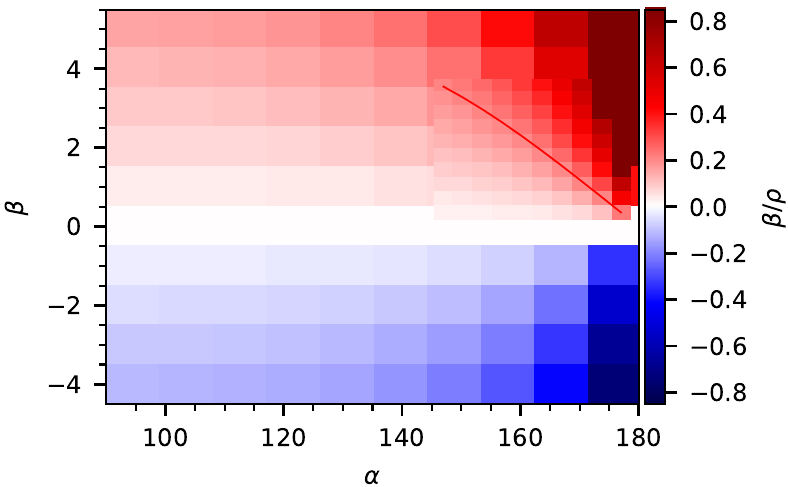}
    \caption{Emission height ({\it the left panel}) and the impact factor divided by the opening angle ({\it the right panel}) for geometries resulting in the pulse width $W \sim 67^{\circ}$. 
    The red line corresponds to the geometries suitable for pulsar \B.
    \label{fig:geometries}
    }
\end{figure*}

\subsection{Geometry}
\label{sec:geometry}

In \cite{2016_Weltevrede} radio polarization measurements were used to constrain the inclination angle $\alpha$ and the impact parameter $\beta$.
The pulse longitude dependence of the observed position angle of the linear polarization can be explained with the rotating vector model \citep{1969_Radhakrishnan, 1970_Komesaroff}.
Taking into account the beam opening angle with constrains on aberration/retardation effects and the observed pulse width can further constrain $\alpha$ and $\beta$ \citep{2015_Rookyard}.
The limitations imposed by the duty cycle of the profile mean, however, that the geometrical parameters are not well constrained.
The geometries acceptable for \B~ range from $\alpha=147^{\circ}$ to $\alpha=177^{\circ}$, and from $\beta=3.5^{\circ}$ to $\beta=0.4^{\circ}$ \citep[see Figure 8 in][]{2016_Weltevrede}, thus in our calculations we use oblique geometries to explain the radio characteristics in \B.
Note that we restrict geometries to the ones consistent with considerations of the expected beam size and with the emission height higher than $2 R_{\rm NS}$.

To investigate variation of the polar-cap electric potential we check the probability of bi-drifting for all the acceptable geometries of PSR \B.
Our aim (and expectation) were to use the presence of the bi-drifting to constrain which geometries were allowed, and thus determine the angles describing the pulsar system.
In order to find bi-drifting we calculated, on the Dutch national supercomputer Cartesius\footnote{\url{https://userinfo.surfsara.nl/systems/cartesius}}, the drift characteristics for 10$^4$ realizations with one crust-anchored magnetic anomaly with random position and strength.
In simulations presented in this paper we use a crust-anchored anomaly with position \mbox{$r_{\rm m,r}=0.95 R_{\rm NS}$}, \mbox{$r_{\rm m, \theta}\in (1^{\circ},70^{\circ})$}, \mbox{$r_{\rm m, \phi} \in (0^{\circ},360^{\circ})$} and strength \mbox{$B_{\rm m}\in (10^{-2}B_{\rm d}, 10^{-1}B_{\rm d})$}, where $R_{\rm NS}$ is the neutron star radius, and \mbox{$B_{\rm d}= 2 \times 10^{12} \,{\rm G}$} is the polar magnetic field strength.
In Table \ref{tab:potential} we show what percentage of realizations exhibit drift characteristics similar to PSR \B~ (with $R<0.1$).
The simulations show that bi-drifting behaviour is most likely when the electric potential variation at the polar cap is described by Equation 3 (i.e. solid-body-like rotation).
However, for all the acceptable geometries the probability is at the similar level ($\sim 1-2 \%$). 
Thus, unfortunately, the presence of bi-drifting does not allow us to further constrain the geometry of \B. 
It appears both methods validate and rule out the same set of geometries, and do not produce the more orthogonal regions in acceptable parameter space.

In order to explore the probability of bi-drifting for a wider range of geometries we consider a 10x10 grid with $\beta \in (-4^{\circ},5^{\circ})$ and $\alpha \in (90^{\circ},180^{\circ})$ as well as a 10x10 grid around the geometries acceptable for \B.
Since the solid-body-like rotation of spark forming regions is the only one which result in the observed phase-locking of pulse components (see Section \ref{sec:phase-locking}), and, moreover, is the one with the highest probability of bi-drifting, hereafter, we describe the global variation at the polar cap with Equation 3.
For every point in the grids we calculate the drift characteristics for 10$^3$ realizations, again with one crust-anchored magnetic anomaly.
Figure \ref{fig:fraction_grid} shows the fraction of realizations showing the bi-drifting characteristics of \B~ (with $R< 0.1$). 
The red line in the figure corresponds to geometries which best fit pulsar \B, as determined from its polarisation profile.
As visible in Figure \ref{fig:fraction_grid}, bi-drifting is to first order more likely for low impact angles $\beta$; but the inclination angle also plays a role. 
To investigate which physical quantity is the actual underlying driver, 
we show in Figure \ref{fig:geometries} (left panel) the emission height chosen to result in the pulse width $W \sim 67^{\circ}$. 
In the right panel we demonstrate the influence of the fractional proximity of the line-of-sight to the pole: the impact factor divided by the opening angle.
The latter Figure shows clear correlation with di-drifting:  for lower values of $\beta / \rho$, \mbox{bi-drifting} is more likely, reaching the maximum probability of $20\%$, when the line of sign cuts the beam centrally ($\beta=0$).
Note again how the red line in Figure \ref{fig:geometries} follows an equal value for $\beta/\rho$.
Since all geometries for \B~ deemed acceptable from its polarisation are thus characterized by a similar value of $\beta/\rho \approx 0.2$, the probability of bi-drifting occurrence cannot be used to narrow down the range of acceptable geometries. 

\begin{figure}[!bt]
  \centerline{\includegraphics[width=7.5cm]{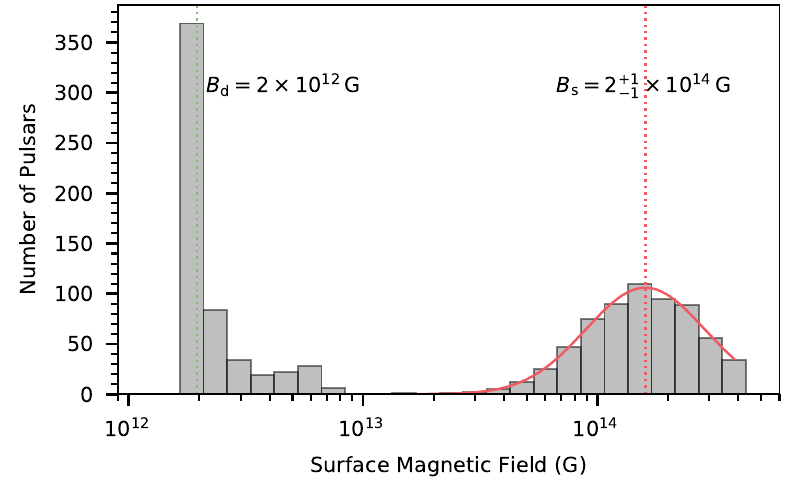}}
    \caption{
    \label{fig:field_hist}
    Surface magnetic field for realizations showing the bi-drifting behaviour (with $R<0.1$).
    }
\end{figure}

\subsection{Conditions at the polar cap}

In this section we present the results of the bi-drifting search and the single pulse modeling for a specific pulsar geometry.
Since the pulsar geometry could not be further constrained from the bi-drifting probability, we use the geometry with the emission height justified by the radio emission models, i.e. with $D\sim 500\,{\rm km}$, \citep[see e.g.][]{1998_Asseo}.
We choose $\alpha=167^{\circ}$ and $\beta=1.5^{\circ}$ with the emission height $D=608\,{\rm km}$ which results in the opening angle $\rho=7.2^{\circ}$ and the pulse width $W=67.1^{\circ}$.
To find the configuration of surface magnetic field which best reproduces the drift characteristics of PSR \B~ we calculate 10$^5$ realizations with one crust-anchored magnetic anomaly.

In  Figure \ref{fig:field_hist} we show a histogram of the surface magnetic field strengths that allow bi-drifting (with \mbox{$R<0.1$}).
It is clear from the histogram that we can distinguish two types of solutions: configurations with a surface magnetic field of the order of the dipolar component, \mbox{$B_{\rm s} \sim B_{\rm d}$}, and configurations with much stronger surface magnetic fields, \mbox{$B_{\rm s} \sim 10^{14} \,{\rm G}$}.
Hereafter we refer to these cases as the weak and strong field configurations.
The bi-drifting is shown by 563 realizations with weak field and 644 realizations with strong field, which corresponds to  $1.2\%$ of all realizations.

\begin{figure}[!bt]
  \centerline{\includegraphics[width=7cm]{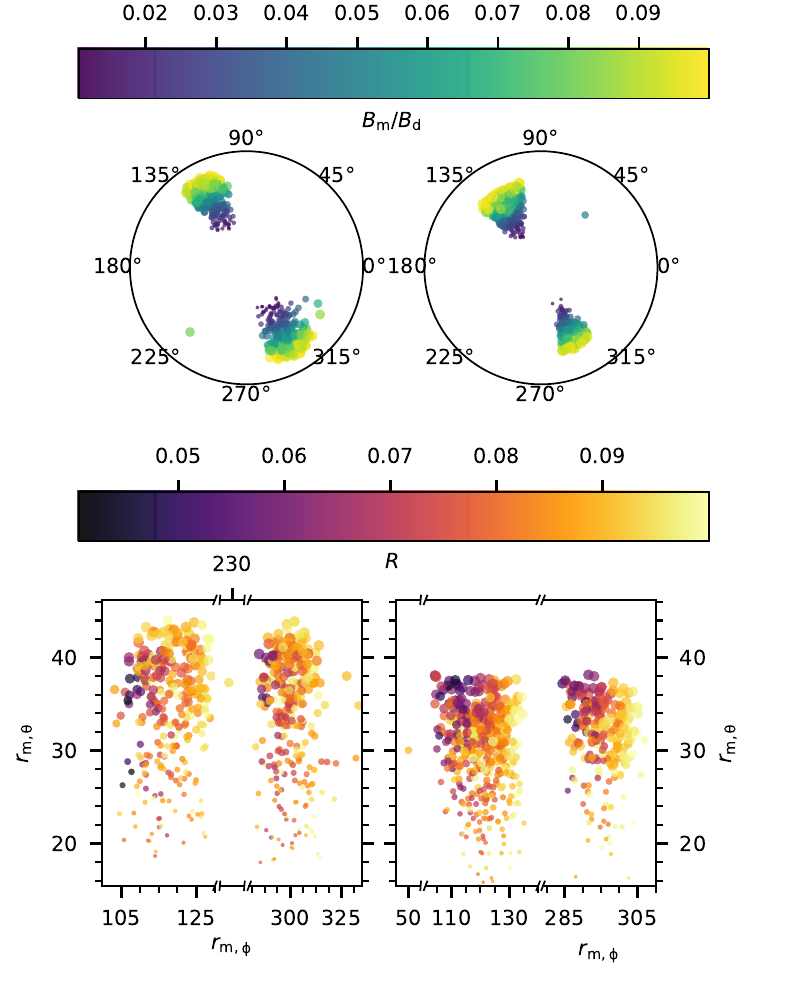}}
    \caption{
    \label{fig:anomaly_locations} 
    Locations of magnetic anomalies for realizations showing bi-drifting behaviour. 
    {\it The left panels} correspond to the weak-field cases, while {\it the right panels} correspond to the strong-field cases.
    The color bars show the anomaly strength ({\it the upper panel}) and the goodness-of-fit ({\it the lower panel}).
    }
\end{figure}

Figure \ref{fig:anomaly_locations} shows the locations of the magnetic anomalies for cases with \mbox{$R<0.1$}.
The weak-field configurations are consequence of anomalies with strength $B_{\rm m}/B_{\rm d}=0.063\pm0.025$ located slightly further from the magnetic axis (with polar angles \mbox{$r_{\rm m, \theta}= 36^{\circ} \pm 6^{\circ}$}) than the anomalies resulting in the strong-field configurations (with \mbox{$r_{\rm m, \theta}= 31^{\circ} \pm 3^{\circ}$} and strength $B_{\rm m}/B_{\rm d}=0.068\pm0.019$).
On the other hand, the azimuthal coordinate of anomalies in both configurations is consistent, \mbox{$r_{\rm m, \phi} = 120^{\circ} \pm 9^{\circ}$} or \mbox{$r_{\rm m, \phi} =297^{\circ} \pm 9^{\circ}$}.
Since we are dealing with a degenerate problem, it is not possible to further constrain the configurations.
However, it is worth to mention, that regardless of the configuration the polar cap is very elongated with eccentricity $e=0.986 \pm 0.1$ for both the weak and strong-field cases.

\begin{figure}[tb]
  \centerline{\includegraphics[width=8cm]{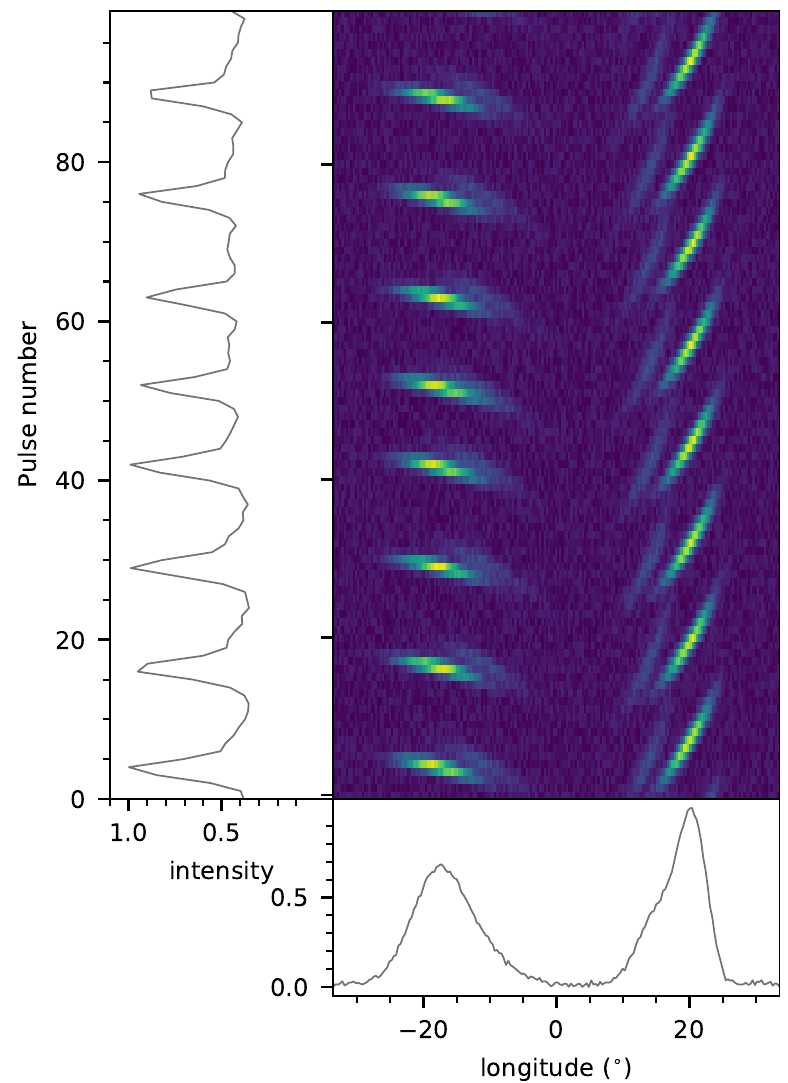}}
    \caption{
    \label{fig:b1839_single}
    Modeled single pulses for the best realization with strong surface magnetic field (with $R=0.07$).
    See Figure \ref{fig:single_pulses} for description of the panels.
    }
\end{figure}

In Figure \ref{fig:b1839_single} we show the modeled single pulses for a realization which best reproduces the drift characteristics of PSR \B.
The surface magnetic field is a result of a magnetic anomaly with strength $B_{\rm m}=0.083 B_{\rm d}$ located at \mbox{($0.95R_{\rm NS}$, $35.05^{\circ}$, $109.21^{\circ}$)} 
that reshapes the magnetic dipole field at the polar cap \mbox{$B_{\rm s} = 1.4 \times 10^{14}\,{\rm G}$}.
The pulses are modeled using $C_{\rm pos}=[0.23, 0.33]$,  $S_{\rm i} = 17$, $S_{\rm spp}=10$, $\sigma_{\rm s}=0.5$ and $P_3=12.3\,{\rm s}$.
To reconstruct the two-component-like integrated profile we assume that emission originating from sparks at the outer track is stronger than from sparks at the inner track with amplitudes $S_{\rm amp}=[1.0, 0.2]$.
Furthermore, we add white noise with amplitude $0.1$ to the signal.

\begin{figure}[tb]
  \includegraphics[width=8.0cm]{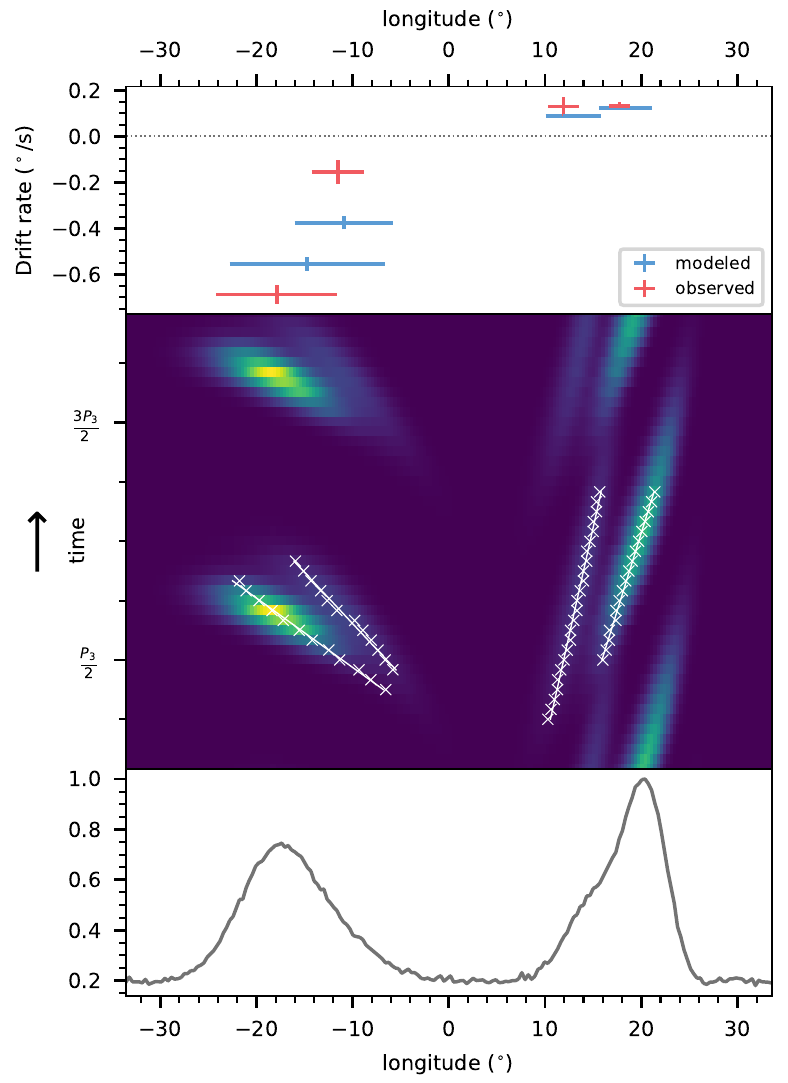}
    \caption{
    Average drift band profile, obtained by folding the single-pulse signal presented in Figure~\ref{fig:b1839_single} with $P_3=12.3 \, P$ (middle panel, shown twice for clarity).
    The top panel shows the measured drift rate, while the bottom panel shows the integrated pulse profile.
    \label{fig:b1839_folded}
    }
\end{figure}

The LRFS analysis on the modeled data results in a frequency $f=0.0811 \pm 0.0003/ P$ which corresponds to $P_3=12.33 \pm 0.04 P$.
In Figure~\ref{fig:b1839_folded} we show the average drift band shape obtained by folding the modeled data with period \mbox{$P_3=12.3 P$}.
We perform similar analysis as for the observed data, i.e. fitting two Gaussians to find subpulse positions, and straight lines to drift band paths to find the modeled drift rates.
The resulting drift rates, $[-0.56 \pm 0.03; \rm -0.38 \pm 0.03; 0.09\pm0.01; 0.12\pm0.01] ^{\circ} / s$, corresponding to the four profile components, match the observed drift rates well, as shown in the top panel of Figure \ref{fig:b1839_folded}.

\section{Discussion} \label{sec:discussion}

The phenomenon of subpulse drifting is expected to be present in at least $40-50 \%$ of the pulsar population.
At present there are around 120 pulsars known to exhibit some form of periodic modulations in their single pulse sequence \citep{2006_Weltevrede, 2007_Weltevrede, 2016_Basu}.
Bi-drifting, an effect where the drift direction of subpulses is different in different pulse profile components, is a rare phenomenon, reported in three radio pulsars, namely J0815+0939, J1034-3224, and B1839-04 \citep{2005_Champion, 2018_Basu, 2006_Weltevrede}.
The varied drift rate observed in different profile components raises a question whether these operate, at some level, independent of each other.
One of the ways to answer this question is to determine if the modulation cycles are phase locked, meaning that if the modulation cycle slows down in one component, it is equally slow in the other component, such that the modulation patterns of both components stay in phase.
In \cite{2016_Weltevrede} it was shown that the magnitude of the observed variability of $P_3$ in \B~ is, within the errors, identical in all components.
The observed phase-locking is, moreover, maintained over a timescale of years.
Such a phase-locking is very common in all pulsars. This suggests that, although very rare, bi-drifting has the same physical origin as regular drifting in other pulsars.

In \cite{2017_Szary} we introduced a modified carousel model to explain bi-drifting behaviour of J0815+0939.
The model is based on the notion that the plasma drift in the inner acceleration region is governed by the global variation of electric potential across the polar cap \citep{2012_Leeuwen}.
In the present paper we explored three different variations of electric potential to explain bi-drifting behaviour of \B: the sinusoidal variation, the variation which leads to keplerian-like rotation of sparks, and the variation which leads to solid-body-like rotation (see Equations 1, 2 and 3).
The drift velocity of spark forming regions depends on the potential variation (see Equation \ref{eq:vdr}), however, the shape of spark track depends on the shape of the actual polar cap rather than the characteristics of potential variation (see Figure \ref{fig:tracks}). 
This, for a given component separation, results in fixed lengths of spark tracks while, on the other hand, the mean velocity of sparks at each track depends on the considered potential.
To reproduce the same value of $P_3$ for each component the number of sparks at each track has to take into account the mean velocity of sparks. 
As a result only a potential variation that leads to solid-body-like rotation ensures the same repetition time of the drift pattern for all components, regardless of their separation (see Section \ref{sec:phase-locking}).
The very strict phase locking of the modulation cycle is a consequence of a quasi-stable structure of a beam pattern and solid-body-like rotation of the spark forming regions.
Furthermore, the potential variation connected with a solid-body-like rotation results in a probability of bi-drifting at least an order magnitude higher than in the other two cases (see Table \ref{tab:potential}).

Note that the above conclusion stands in opposition to equidistant sparks distribution proposed by \cite{2000_Gil}. More specifically, in that model, the number of sparks along the inner track would naturally be less than those along the outer track. A rigid, in-phase rotation of two such carousels cannot result in the observed phase-locking between different components. In fact, this realization was one of the primary motivating factors for \citet{2014_MD} to propose the same-frequency multi-altitude scenario \citep{1993_Rankin} as an explanation for the observed phase-locking, particularly in PSR~B1237+25, and generally in all multi-component pulsars. More recently, \citet{2019_Maan} showed that within the above multi-altitude scenario, the leading and trailing component pairs are expected to exhibit phase offsets of the same magnitude, but in opposite directions. 
We note that our model would in general result in different values of phase offset for the leading and trailing components, however, sign-reversed offsets with equal magnitude can also be realized for selected viewing geometry parameters and number of sparks. Measurements of the phase-offsets in other pulsars with multiple components would help in clarifying the actual underlying cause of the observed phase-locking.

The observed position angle of the linear polarization can be used to constrain the magnetic axis inclination angle $\alpha$ and the impact parameter $\beta$ using the rotating vector model \citep{1969_Radhakrishnan, 1970_Komesaroff}.
However, the method is limited due to the duty cycle of the pulsar.
In this paper, we have explored the influence of the pulsar geometry on the observed drift characteristics.
We have shown that bi-drifting behaviour is more likely for central cuts of the emission beam, reaching a maximum probability of $20\%$ for $\beta = 0$ (see Figure \ref{fig:fraction_grid}). 
The parameter that affects the bi-drifting probability is $\beta / \rho$, which describes how close the line of sight approaches the magnetic axis.
Since all geometries for \B~ acceptable from polarisation studies are characterised with similar $\beta / \rho \sim 0.2$, the bi-drifting probability cannot be used to further constrain pulsar geometry.

In this paper we aimed to reproduce the bi-drifting behaviour of \B.
The non-dipolar nature of our best-fitting realizations produces the elliptic, tilted spark trajectories proposed in \cite{2017_Wright}.
We have modeled a large sample of $10^5$ possible configurations resulting in 1207 realizations showing bi-drifting.
The strength of surface magnetic field falls into one of two possible scenarios, corresponding to a relatively weak ($\sim 10^{12} \,{\rm G}$) or strong ($\sim 10^{14} \,{\rm G}$) field.
The resemblance with the bifurcation of solutions for \J~ (see Figure 15 in \citealt{2017_Szary}) may be due to the fact that for both pulsars we used magnetic anomalies of a similar nature: crust-anchored (about $500$ meters below the surface), with strengths \mbox{$B_{\rm m}\in (10^{-2}B_{\rm d}, 10^{-1}B_{\rm d})$}.

\section{Conclusions} \label{sec:conclusions}
Based on the drift properties of PSR \B~ we were able to constrain how the electric potential varies across the polar cap. 
A fundamental problem of the observed phase-locking of components is resolved by concluding the electric potential must lead to solid-body-like rotation of spark forming regions.
Furthermore, we find that solid-body-like rotation is the preferred type of rotation to explain bi-drifting in \B.
Although bi-drifting is rare, this trait did not translate into a specific solution for the geometry of \B. 
We did find that bi-drifting is more likely for sight-line cuts close to the magnetic pole (i.e. lower values of $\beta / \rho$).

\acknowledgments

We would like to thank the anonymous referee, whose constructive comments improved clarity of the paper in a significant way.

This research received funding from the Netherlands Organisation for Scientific Research (NWO) under project "CleanMachine" (614.001.301), from the European Research Council under the European Union’s Seventh Framework Programme (FP/2007-2013) / ERC Grant Agreement n. 617199.

Part of this work was carried out on the Dutch national e-infrastructure with the support of SURF Cooperative. The research was supported by a NWO Science subsidy for the use of the National Computer Facilities (project 16675).


\appendix 
 
\section{The modified carousel model}
\label{app}

\subsection{Origin of the carousel model}

\cite{1975_Ruderman} considered discharges of small regions over the polar cap, the so-called sparks. 
Since plasma filaments injected into the magnetosphere result in the generation of coherent radio emission, it seems reasonable that the location of the sparks at the polar cap determines the location of subpulses within a pulsar's pulse profile.
\cite{1975_Ruderman} noticed that periods of spark circulation around the magnetic axis in a calculable way seem to match the observed periods of drifting subpulses (see Appendix \ref{app:modifications}).
As a result, the carousel model was introduced which assumes carousel-like rotation of sparks around the magnetic axis.
Note that such model stands in opposition to the interpretation that drift is simply due to plasma lagging behind the co-rotation, as for an inclined rotator (the geometry considered in \citealt{1975_Ruderman}) the carousel model requires that in parts of the polar cap sparks move faster than co-rotating plasma.

\subsection{Modifications of the carousel model}
\label{app:modifications}

Progress in the analysis of drifting subpulses revealed that the circulation speed in a pure vacuum gap is too high when compared with observations.
Moreover, some pulsars  demonstrate time variations in the drift rate, including a change of the apparent drift direction, which seems to be inconsistent with the carousel model.
\cite{2012_Leeuwen} pointed out that the drift velocity depends not on the absolute value, but on the variation of the accelerating potential across the polar cap.
This idea was further developed in \cite{2017_Szary}, where it was shown that sparks do not rotate around the magnetic axis per se, but rather around the point of electric potential extremum at the polar cap: minimum in the pulsar case ($\mathbf{\Omega} \cdot \mathbf{B} < 0$) and maximum in the antipulsar case ($\mathbf{\Omega} \cdot \mathbf{B} > 0$).
Since radio emission is generated at altitudes where magnetic field is purely dipolar, whenever the electric potential extremum is in the center of the polar cap, the resulting subpulses should rotate around the magnetic axis. 
 
In the paper at hand we consider pulsar geometries with $\mathbf{\Omega} \cdot \mathbf{B} < 0$, where sparks form in regions of local potential minima at the polar cap.

\subsection{Physics of the modified carousel model}

In general in pulsar magnetosphere there are two sources of electric field: an inductive electric field, due to time-changing magnetic field as pulsar rotates and a potential field, due to the charge density in the magnetosphere.

\subsubsection{Anti-aligned rotator}

\begin{figure}[tb]
    \centering
    \includegraphics[width=7.0cm]{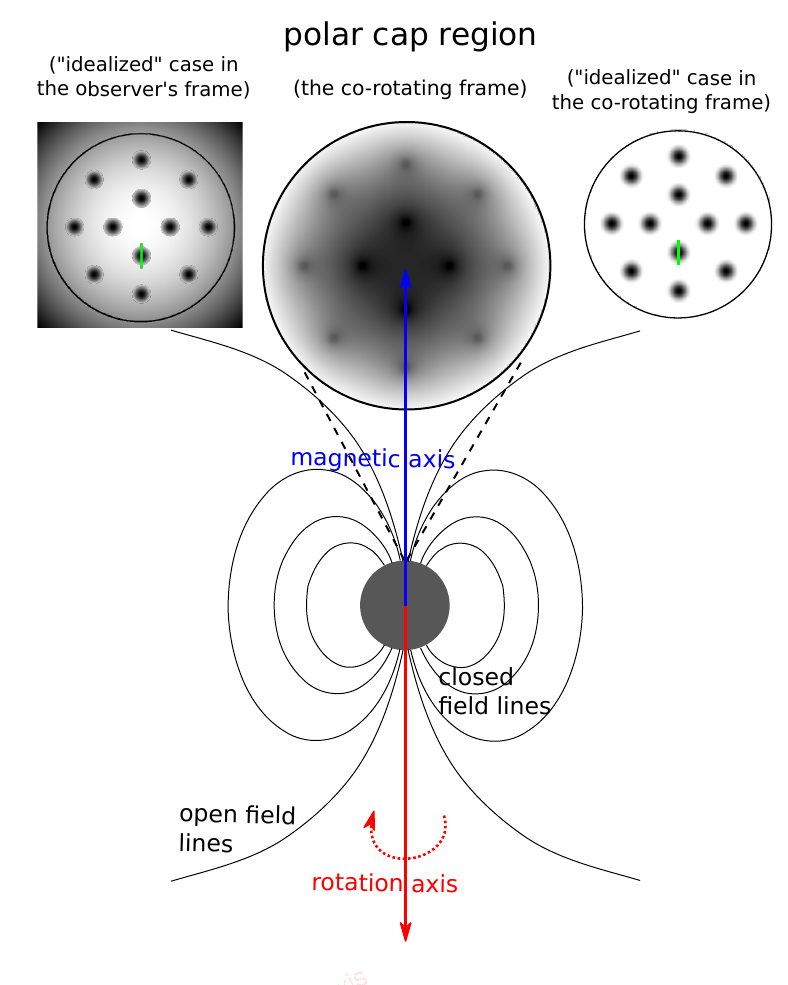}
    \caption{
    Electric potential in the polar cap region for an anti-aligned geometry. The left panel corresponds to the observer's frame of reference, while the center and right panels correspond to the co-rotating frame of reference. The electric potential in the left and right panels are shown for an "idealized" case with no particle acceleration along magnetic field lines ($E_\parallel=0$) in regions between sparks. Darker level of grey corresponds to a lower electric potential. The green line marks a region considered in Figure \ref{fig:spark}.
    \label{fig:anti-align}
    }
\end{figure}

In the case when the magnetic and rotation axes are anti-aligned (with $\alpha=180^\circ$), the inductive electric field is zero and there is no magnetic dipole radiation. 
In Figure \ref{fig:anti-align} we show electric potential in the polar cap region for an anti-aligned rotator.
We consider two cases: an "idealized" case with no acceleration along magnetic field lines ($E_\parallel=0$) in regions between sparks (the left and right panels), and a case when lack of plasma in spark formation region is not fully compensated in regions between sparks resulting in $E_\parallel \neq 0$ (the center panel). 

\begin{figure}[tb]
    \centering
    \includegraphics[width=12.0cm]{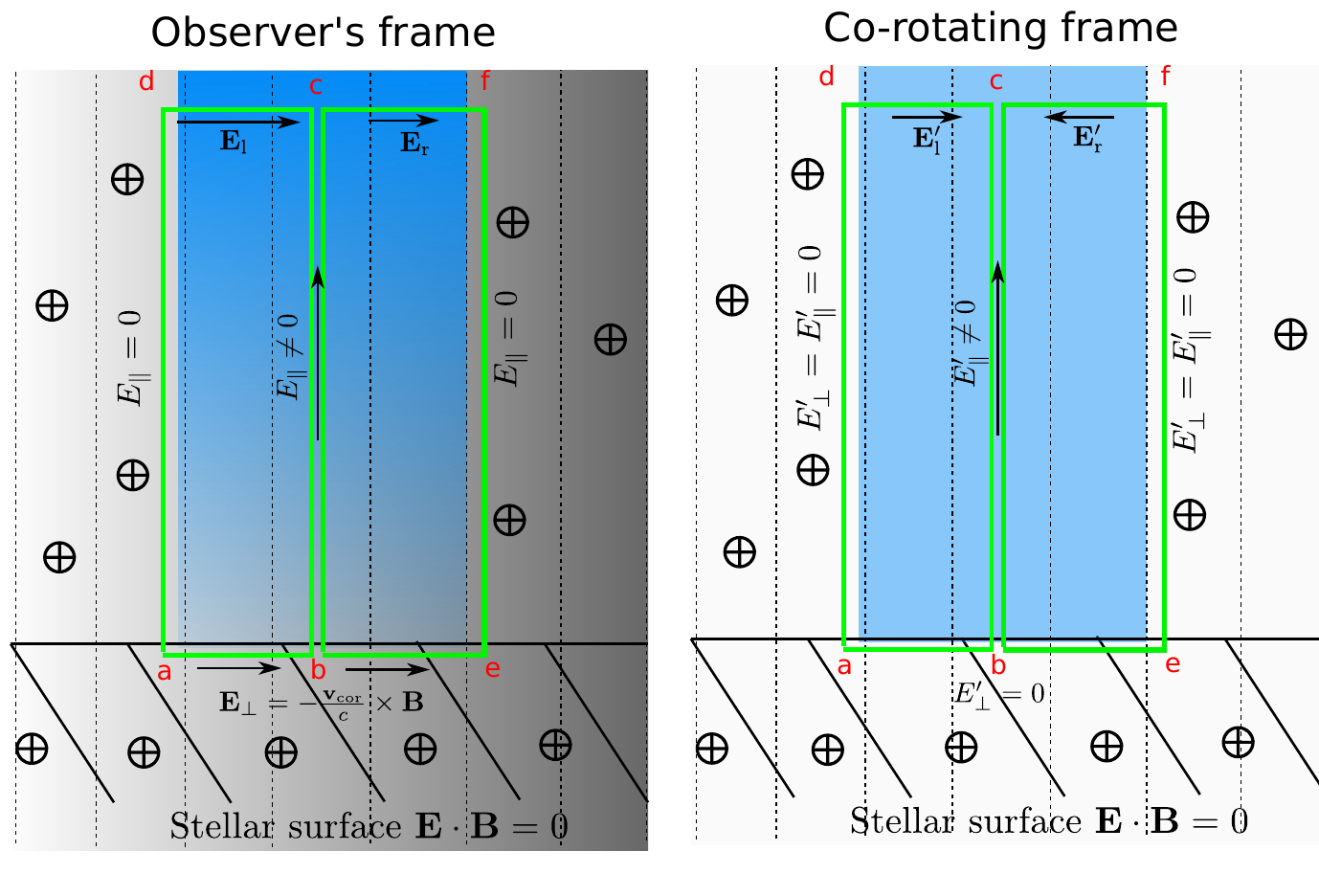}
    \caption{
    Electric field within a region of spark formation marked with the green line in Figure \ref{fig:anti-align} for an idealized case (with no accelerating electric field in regions between sparks). The left panel corresponds to the observer's frame while the right panel corresponds to the co-rotating frame of reference. Darker level of grey and blue in the left panel correspond to a lower potential value.
    \label{fig:spark}
    }
\end{figure}

To find the drift direction in the "idealized" case, in Figure \ref{fig:spark} we consider electric field within a region of spark formation (see the green line in Figure \ref{fig:anti-align}).
Since for an anti-aligned rotator $\nabla \times {\bf E} = 0$ with high accuracy \citep[see, e.g.,][]{2012_Leeuwen}, the circulation of the electric field along a closed path is zero.
Let us consider two paths $a b c d$ and $b e f c$. 
The bottom parts ($a b$ and $b e$) are just below the neutron star surface, which we assume to be a perfect conductor, while the top parts ($c d$ and $f c$) cross the acceleration gap; the lateral sides ($d a$ and $e f$) follow magnetic field lines in the regions beyond a spark. 
Note that here we consider the idealized case where there is no acceleration along magnetic field lines outside of a spark ($E_\parallel=0$).
The circulation of the electric field along the paths is

\begin{equation}
\begin{split}
    \oint {\bf E \cdot dl}  & = \int_a^b {\bf E_\perp \cdot dx} + \int_b^c {\bf E_\parallel \cdot dy} + \int_c^d {\bf E_l \cdot dx}  = 0 \hspace{1cm} \int_c^{d} {\bf E_l \cdot dx} = \Delta V_{a b} + \Delta V_{b c} = (V_b - V_{a}) +  (V_c - V_b)\\ 
    \oint {\bf E \cdot dl}  & = \int_{b}^e {\bf E_\perp \cdot dx} + \int_f^c {\bf E_r \cdot dx} + \int_c^{b} {\bf E_\parallel \cdot dy}  = 0  \hspace{1cm} \int_f^{c} {\bf E_r \cdot dx} = \Delta V_{b e} + \Delta V_{c b} = (V_e - V_{b}) + (V_b - V_c), 
\end{split}
\end{equation}
where $E_\perp$ and $E_\parallel$ are components of electric field, perpendicular and parallel to the magnetic field, respectively, while $\Delta V_{i j} = V_j - V_i$ is the electric potential difference from point $i$ to $j$. 
Since $V_a > V_{b}$ and $V_b > V_c$, direction of vector $\bf E_l$ is opposite to $\overrightarrow{c d}$.
Furthermore, since $V_e < V_b$ and $| \Delta V_{b e}| > |\Delta V_{c b}|$ (path inside the acceleration region), $\bf E_r$ is opposite to $\overrightarrow{fc}$, and $E_r < E_\perp < E_l$ (see the left panel in Figure \ref{fig:spark}).
To study the drift of plasma within a spark with respect to the co-rotating magnetosphere, it is convenient to use the co-rotating frame of reference.
Therefore, due to the Lorentz transformation, ${\bf E}^\prime = {\bf E} + c^{-1} ({\bf \Omega} \times {\bf r})  \times {\bf B} = {\bf E} - {\bf E}_\perp$, here $E$ is the electric field in the observer's frame, while $E^{\prime}$ is the electric field in the co-rotating frame.

As a result, electric field in the co-rotating frame is directed towards the center of a spark (see the right panel in Figure \ref{fig:spark}).
Such a configuration indicates rotation of plasma around the center of a spark forming region \citep[see, e.g.,][]{2010_Kontorovich}, resulting in no systematic drift of plasma with respect to the co-rotating magnetosphere.

On the other hand, if we assume that lack of plasma in spark forming regions is not fully compensated in regions between sparks by plasma generation or charge flow (the center panel in Figure \ref{fig:anti-align}), the charged particles in those regions exhibit systematic drift around the potential minimum at the polar cap (see Figure 8 in \citealt{2017_Szary}).
In other words, in the modified carousel model the polar cap consist of spark forming regions, where plasma responsible for radio emission is produced, and regions between sparks, where charged particles drift around the electric potential extremum at the polar cap.
The observed stability of subpulse structures suggest that the pattern of discharging regions is also stable.
Although, the exact mechanism of formation of the discharging pattern at the polar cap is still unknown, from the considerations presented above we can draw a few conclusions.
First, since the model requires that the accelerating electric field ($E_\parallel$) in regions between sparks is not fully screened, we expect acceleration of charged particles in those regions.
Second, there must be a mechanism which prevents or limits discharge between sparks such that the plasma in those regions does not generate radio emission or this radiation is significantly lower in intensity.
One of the possible scenario is reverse plasma flow induced by a mismatch between the magnetospheric current distribution and the current injected in the spark forming regions \citep[see, e.g., ][]{2012_Lyubarsky}.
However, to get a full picture of formation of discharging regions at the polar cap a 3D particle-in-cell simulation is required.

\subsubsection{Oblique rotator}

In this work, we use an oblique geometry ($\mathbf{\Omega} \cdot \mathbf{B} < 0$) to explain observed radio emission characteristics in \mbox{B1839-04}. 
While the electrodynamics of such an oblique rotator is still an unsolved problem \citep{2016_Melrose}, 
we can use the analogy to an anti-aligned rotator to understand the drifting phenomenon in a general case.

As the pulsar rotates, the time changing magnetic field gives rise to an inductive electric field, $E_{\rm ind}$, which can be estimated using the vacuum dipole model \citep{1955_Deutsch}.
On the other hand, \citet{1969_Goldreich} have shown that "a rotating magnetic neutron star cannot be surrounded by a vacuum" introducing the rotating magnetosphere.
In the magnetosphere the co-rotation charge density gives rise to a potential (curl-free) electric field, $E_{\rm pot}$.
As a result, the co-rotation electric field has both inductive and potential components, $E_{\rm cor} = E_{\rm ind} + E_{\rm pot}$.
Thereby, the plasma velocity in the oblique co-rotation model may be regarded as the sum of the drift velocities due to $E_{\rm ind}$ and $E_{\rm pot}$.  
In the spark forming regions, the lack of plasma affects only the potential component of electric field, $E_{\rm pot}$.
Thus, to estimate the behaviour of charges in the polar cap region with respect to the rotating plasma (i.e. in the co-rotating frame of reference),
we can use the same reasoning as in the anti-aligned rotator (see the previous Section).

To summarize, the modified carousel model requires that the temporary lack of plasma in spark forming regions is not fully compensated in regions between sparks, giving a rise to a global variation of electric potential at the polar cap. 
For the pulsar geometry ($\mathbf{\Omega} \cdot \mathbf{B} < 0$), the charged particles between sparks rotate around the electric potential minimum at the polar cap.
The radio emission is generated at much higher altitudes ($~\sim 500 \, {\rm km}$); if the potential minimum coincides with the center of the polar cap,  subpulses drift around the magnetic axis even for a non-dipolar configuration of surface magnetic field.


\bibliographystyle{aasjournal}

\begin{thebibliography}{}

\bibitem[\protect\citeauthoryear{Asseo \& Melikidze}{1998}]{1998_Asseo} Asseo E., Melikidze G.~I., 1998, MNRAS, 301, 59

\bibitem[{{Backer}(1970)}]{1970_Backer} {Backer}, D.~C. 1970, \nat, 227, 692

\bibitem[Basu et al.(2016)]{2016_Basu} Basu, R., Mitra, D., Melikidze, G.~I., et al.\ 2016, \apj, 833, 29 

\bibitem[Basu, \& Mitra(2018)]{2018_Basu} Basu, R., \& Mitra, D.\ 2018, \mnras, 475, 5098.

\bibitem[Basu et al.(2019)]{2019_Basu} Basu, R., Mitra, D., Melikidze, G.~I., et al.\ 2019, \mnras, 482, 3757.

\bibitem[{{Champion} {et~al.}(2005){Champion}, {Lorimer}, {McLaughlin},
  {Xilouris}, {Arzoumanian}, {Freire}, {Lommen}, {Cordes}, \&
  {Camilo}}]{2005_Champion}
{Champion}, D.~J., {Lorimer}, D.~R., {McLaughlin}, M.~A., {et~al.} 2005,
  \mnras, 363, 929

\bibitem[Chen \& Ruderman(1993)]{1993_Chen} Chen, K., \& Ruderman, M.\ 1993, \apj, 402, 264

\bibitem[Clifton \& Lyne(1986)]{1986_Clifton} Clifton, T.~R., \& Lyne, A.~G.\ 1986, \nat, 320, 43 

\bibitem[Deutsch(1955)]{1955_Deutsch} Deutsch, A.~J.\ 1955, Annales d'Astrophysique, 18, 1

\bibitem[Drake \& Craft(1968)]{1968_Drake} Drake, F.~D., \& Craft, H.~D.\ 1968, Nature, 220, 231 

\bibitem[{{Gil} \& {Sendyk}(2000)}]{2000_Gil} {Gil}, J.~A., \& {Sendyk}, M. 2000, \apj, 541, 351

\bibitem[{{Gil} {et~al.}(2002){Gil}, {Melikidze}, \& {Mitra}}]{2002_Gil} {Gil}, J.~A., {Melikidze}, G.~I., \& {Mitra}, D. 2002, \aap, 388, 235

\bibitem[Goldreich \& Julian(1969)]{1969_Goldreich} Goldreich, P., \& Julian, W.~H.\ 1969, \apj, 157, 869

\bibitem[Hobbs et al.(2004)]{2004_Hobbs} Hobbs, G., Lyne, A.~G., Kramer, M., Martin, C.~E., \& Jordan, C.\ 2004, \mnras, 353, 1311 

\bibitem[Kijak \& Gil(2003)]{2003_Kijak} Kijak, J., \& Gil, J.\ 2003, \aap, 397, 969 

\bibitem[Komesaroff(1970)]{1970_Komesaroff} Komesaroff, M.~M.\ 1970, \nat, 225, 612 

\bibitem[Kontorovich(2010)]{2010_Kontorovich} Kontorovich, V.~M.\ 2010, Soviet Journal of Experimental and Theoretical Physics, 110, 966

\bibitem[Lyubarsky(2012)]{2012_Lyubarsky} Lyubarsky, Y.\ 2012, \apss, 342, 79

\bibitem[Maan \& Deshpande(2014)]{2014_MD} Maan, Y., \& Deshpande, A.~A.\ 2014, \apj, 792, 130

\bibitem[Maan(2019)]{2019_Maan} Maan, Y.\ 2019, \apj, 870, 110

\bibitem[Melikidze et al.(2000)]{2000_Melikidze} Melikidze, G.~I., Gil, J.~A., \& Pataraya, A.~D.\ 2000, \apj, 544, 1081 

\bibitem[Melrose(1967)]{1967_Melrose} Melrose, D.~B.\ 1967, \planss, 15, 381

\bibitem[Melrose \& Yuen(2016)]{2016_Melrose} Melrose, D.~B., \& Yuen, R.\ 2016, Journal of Plasma Physics, 82, 635820202

\bibitem[Mitra \& Rankin(2002)]{2002_Mitra} Mitra, D., \& Rankin, J.~M.\ 2002, \apj, 577, 322 

\bibitem[Pacini(1967)]{1967_Pacini} Pacini, F.\ 1967, \nat, 216, 567

\bibitem[Pacini(1968)]{1968_Pacini} Pacini, F.\ 1968, \nat, 219, 145

\bibitem[Radhakrishnan \& Cooke(1969)]{1969_Radhakrishnan} Radhakrishnan, V., \& Cooke, D.~J.\ 1969, \aplett, 3, 225 

\bibitem[Rankin(1993)]{1993_Rankin} Rankin, J.~M.\ 1993, \apj, 405, 285

\bibitem[Rankin \& Rosen(2014)]{2014_Rankin} Rankin, J., \& Rosen, R.\ 2014, \mnras, 439, 3860 

\bibitem[Rookyard et al.(2015)]{2015_Rookyard} Rookyard, S.~C., Weltevrede, P., \& Johnston, S.\ 2015, \mnras, 446, 3367 

\bibitem[Ruderman \& Sutherland(1975)]{1975_Ruderman} Ruderman, M.~A., \& Sutherland, P.~G.\ 1975, \apj, 196, 51 

\bibitem[Serylak et al.(2009)]{2009_Serylak} Serylak, M., Stappers, B.~W., \& Weltevrede, P.\ 2009, \aap, 506, 865 

\bibitem[{{Szary} {et~al.}(2015){Szary}, {Melikidze}, \& {Gil}}]{2015_Szary}
{Szary}, A., {Melikidze}, G.~I., \& {Gil}, J. 2015, \mnras, 447, 2295

\bibitem[Szary \& van Leeuwen(2017)]{2017_Szary} Szary, A., \& van Leeuwen, J.\ 2017, \apj, 845, 95 

\bibitem[Szary, \& van Leeuwen(2019)]{2019_Szary} Szary, A., \& van Leeuwen, J.\ 2019, \apj, 878, 163

\bibitem[van Leeuwen \& Timokhin(2012)]{2012_Leeuwen} van Leeuwen, J., \& Timokhin, A.~N.\ 2012, \apj, 752, 155 

\bibitem[{{Weltevrede} {et~al.}(2006){Weltevrede}, {Edwards}, \&
  {Stappers}}]{2006_Weltevrede}
{Weltevrede}, P., {Edwards}, R.~T., \& {Stappers}, B.~W. 2006, \aap, 445, 243

\bibitem[Weltevrede et al.(2007)]{2007_Weltevrede} Weltevrede, P., Stappers, B.~W., \& Edwards, R.~T.\ 2007, \aap, 469, 607 

\bibitem[Weltevrede(2016)]{2016_Weltevrede} Weltevrede, P.\ 2016, \aap, 590, A109 

\bibitem[Wright \& Weltevrede(2017)]{2017_Wright} Wright, G., \& Weltevrede, P.\ 2017, \mnras, 464, 2597 


\end{thebibliography}

\end{document}